\documentclass[slac_one]{revtex4}
\usepackage{graphicx}
\usepackage{fancyhdr}
\newcommand{\pt}{\ensuremath{p_{T}}}
\newcommand{\st}{\ensuremath{S_{T}}}
\newcommand{\met}{\ensuremath{E\!\!\!\!/_T}}

\newcommand{\pbarn} {\ensuremath{\mathrm{pb^{-1}}}}
\newcommand{\fbarn} {\ensuremath{\mathrm{fb^{-1}}}}

\begin{document}

\title{{\small{The 34th International Conference on High Energy Physics, Philadelphia, Pennsylvania, USA}}\\ 
\vspace{12pt}
Search for Leptoquarks, Excited Leptons and Technicolor at the LHC} 

%

\author{Vikas Bansal (ATLAS Collaboration)}
\affiliation{University of Pittsburgh, 3941 O'Hara St., Pittsburgh, PA~15260, USA}

\begin{abstract}
The ATLAS and CMS experiments at the Large Hadron Collider (LHC) will soon
search for physics phenomena that are not predicted by the Standard Model.
Technicolor, Compositeness and GUT-based models are rich in high-$\pt$ leptons
and could be studied in such final states. This contribution shows studies that indicate that a 5 $\sigma$ discovery of a 500 GeV leptoquark could occur with 100 $\pbarn$ of integrated luminosity, Technicolor predicted particles could be seen with 4 $\fbarn$ of integrated luminosity and, excited electron could be detected with 300 $\fbarn$ of integrated luminosity. 
\end{abstract}

\maketitle

\thispagestyle{fancy}


\section{INTRODUCTION} 
The Large Hadron Collider (LHC) will soon open up a new energy scale that will directly probe for physical phenomena outside the framework of the Standard Model (SM). 
Many Grand Unification inspired extensions of the SM introduce new, very heavy particles such as leptoquarks. The finite number of fermion generations in the SM could be explained by composite models thereby giving rise to excited fermions such as excited electrons. The mechanism of electroweak symmetry breaking (EWSB) in the SM is based on a purely weakly-interacting Higgs sector in which, the mass of the Higgs boson, itself, is unstable because of quadratic divergences due to loop corrections. This mechanism could as well be explained in the context of strong dynamics, e.g. Technicolor model.

In this work, we discuss the discovery potential of ATLAS and CMS experiments at the LHC to these new particles.


\section{SEARCH FOR SCALAR LEPTOQUARKS WITH ATLAS}

\begin{figure}[htbp]
\center{
{\includegraphics[width=0.4\textwidth]{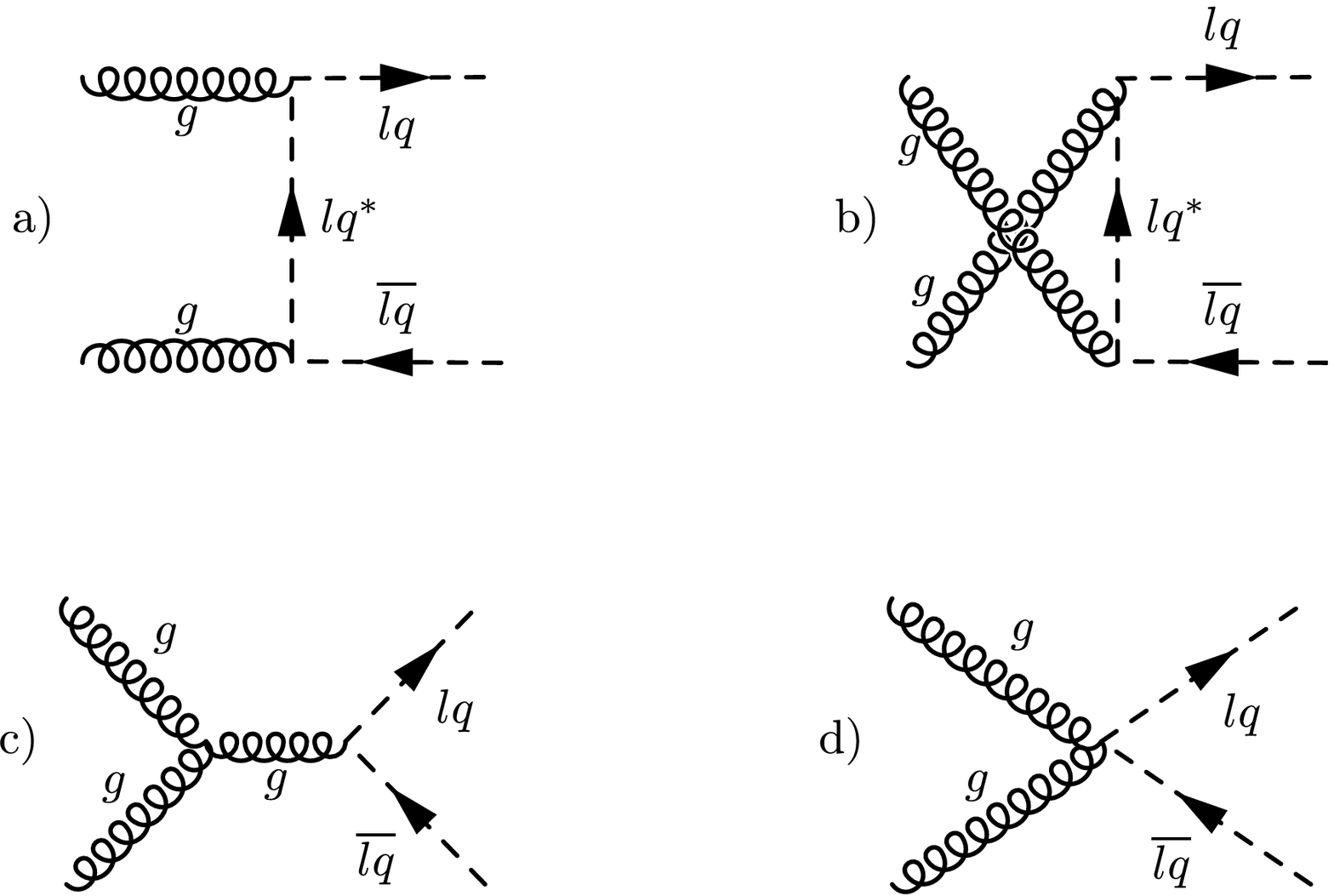}}   
\hspace{0.75in}
{\includegraphics[width=0.4\textwidth]{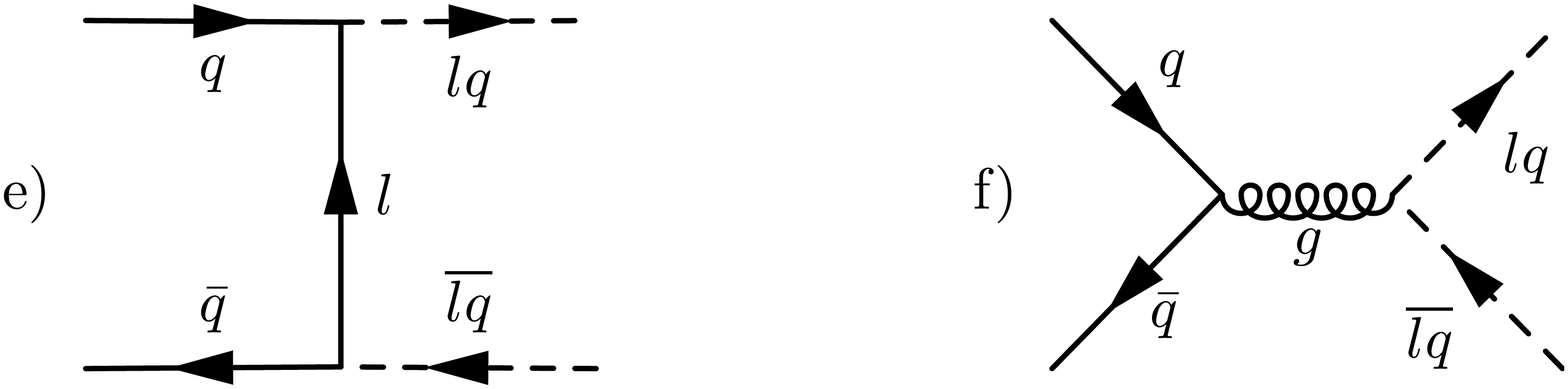}}
}
\caption{\small {
Feynman diagrams for pair production of scalar leptoquarks via
	gluon-gluon fusion (a-d) and quark anti-quark annihilation (e-f). 
}}
\label{fig:LQ_feynman}
\end{figure}

Leptoquarks (LQ) are hypothetical bosons carrying both quark and lepton quantum numbers, as well as fractional electric charge\cite{Pati:1974yy,Eichten:1986eq,Eichten:1983hw,Buchmuller:1986iq,Georgi:1974sy}.  
Leptoquarks could, in principle, decay into any combination of any flavor lepton and any flavor quark. 
Experimental limits on lepton number violation, 
flavor-changing neutral currents, and proton decay 
favor three generations of leptoquarks. 
In this scenario, each leptoquark couples to a lepton 
from the same SM generation and a quark\cite{Leurer:1993em}. 
Leptoquarks can either be produced in pairs by the strong interaction or
in association with a lepton via the leptoquark-quark-lepton coupling. Figure \ref{fig:LQ_feynman} shows Feynman diagrams for pair production of leptoquarks at the LHC.

This contribution describes the search strategy for 
leptoquarks decaying to either an electron and a quark or a muon and 
a quark leading to final states with two leptons and at least two jets. The branching fraction of a leptoquark to a charged lepton and a quark is denoted as $\beta$\footnote{$\beta = 1$ would mean that leptoquarks do not decay into quarks and charge-neutral leptons, neutrinos.}.

The signal has been studied\cite{ATLAS_CSC_PHYSICS_BOOK:2008} using Monte Carlo (MC) samples for first generation (1st~gen.) 
and second generation (2nd~gen.) scalar leptoquarks simulated 
at four masses of 300~GeV, 400~GeV, 600~GeV, and 800~GeV 
with the MC generator {\sc Pythia} \cite{pythia} at 14~TeV $pp$ center-of-mass energy.
The next to leading order (NLO) cross section\cite{Kramer:2004df} for the above simulated signal decreases with leptoquark mass from a few pb to a few fb with mass point of 400~GeV at (2.24$\pm$0.38)~pb.

Signal reconstruction requires selection of two high quality leptons and at least two jets. Each signal jet and lepton candidate is required to have $\pt>$ 20~GeV. This helps to suppress low $\pt$ background predicted by the SM. Leptons are required to have pseudorapidity $|\eta|$ below 2.5, which is the inner detector's acceptance, whereas jets are restricted to $|\eta| < 4.5$ to suppress backgrounds from underlying event and minimum bias events that dominate in the forward region of the detector. In addition, leptons are required to pass identification criteria, which, in case of electrons, are based on electromagnetic-shower shape variables in the calorimeter and, in the case of muons, are based on finding a common track in the muon spectrometer and the inner detector together with muon isolation\footnote{$E_T^{iso}/p_T^\mu \le 0.3$, where 
         $p_T$ is muon candidate's transverse momentum and $E_T^{iso}$ is energy detected in the calorimeters 
         in a cone of $\Delta$R=$\sqrt{(\Delta\eta^2 + \Delta\phi^2)}$=0.2 around muon candidate's reconstructed trajectory.} requirement in the calorimeter. Electron candidates are also required to have a matching track in the inner detector. Furthermore, it is also required that signal jet candidates are spatially separated from energy clusters in the electromagnetic calorimeter that satisfy electron identification criteria. Finally, a pair of leptoquark candidates are reconstructed from lepton-jet combinations. Given the fact that these four objects can be combined to give two pairs, the pair that has minimum mass difference between the two leptoquark candidates is assumed to be the signal.

The main backgrounds to the signal come from $t\bar{t}$, and $Z /DY$+jets production processes. Multijet production where two jets are misidentified as electrons, represents another background to dielectron(1st~gen.)~channel. 
In addition, minor contributions arise from diboson production. 
Other potential background sources, such as single-top production, were also studied and found to be insignificant.  

\begin{figure}[htbp]
\center{
{\includegraphics[width=2.4in]{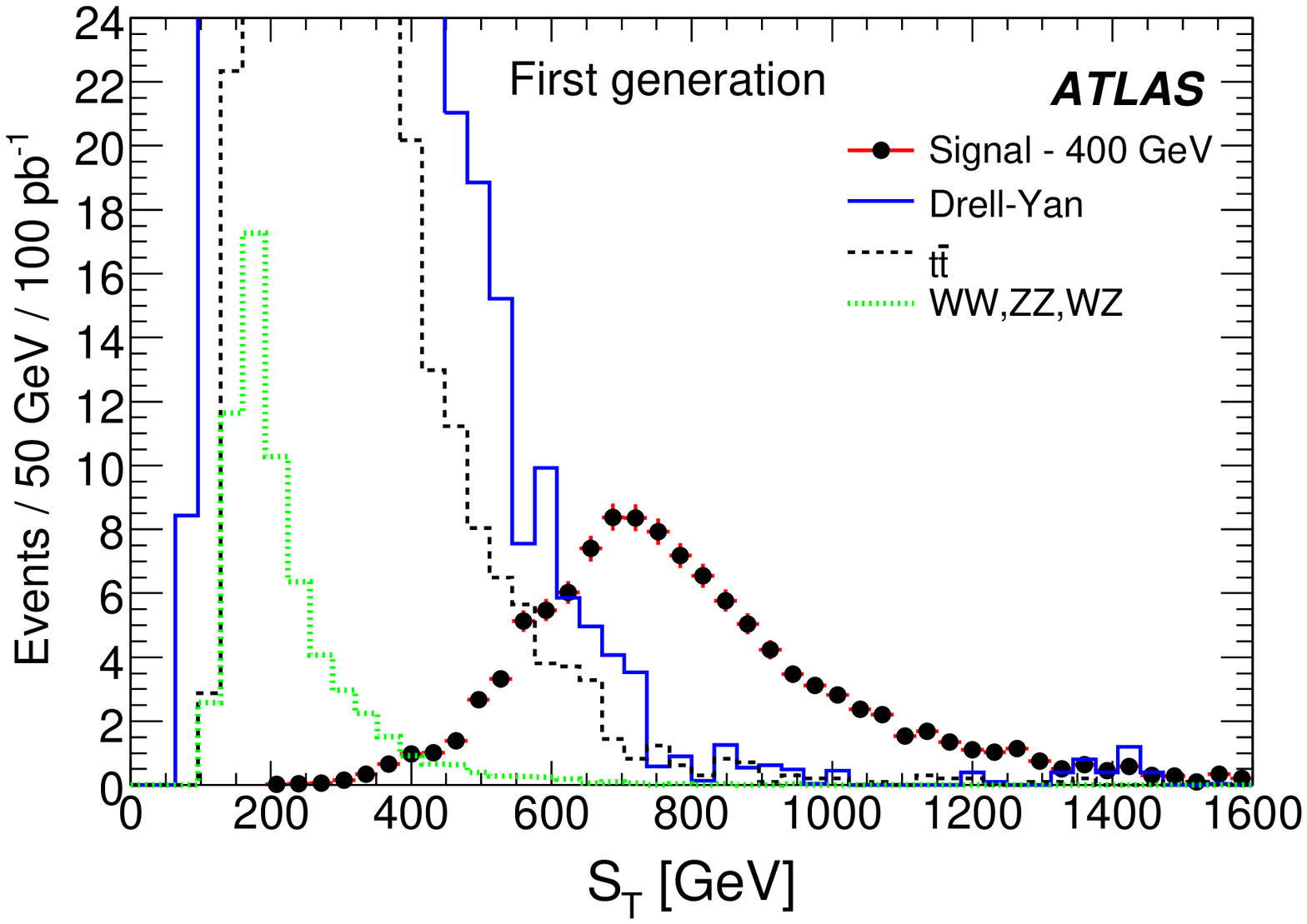}}
\hspace{0.5in}
{\includegraphics[width=2.4in]{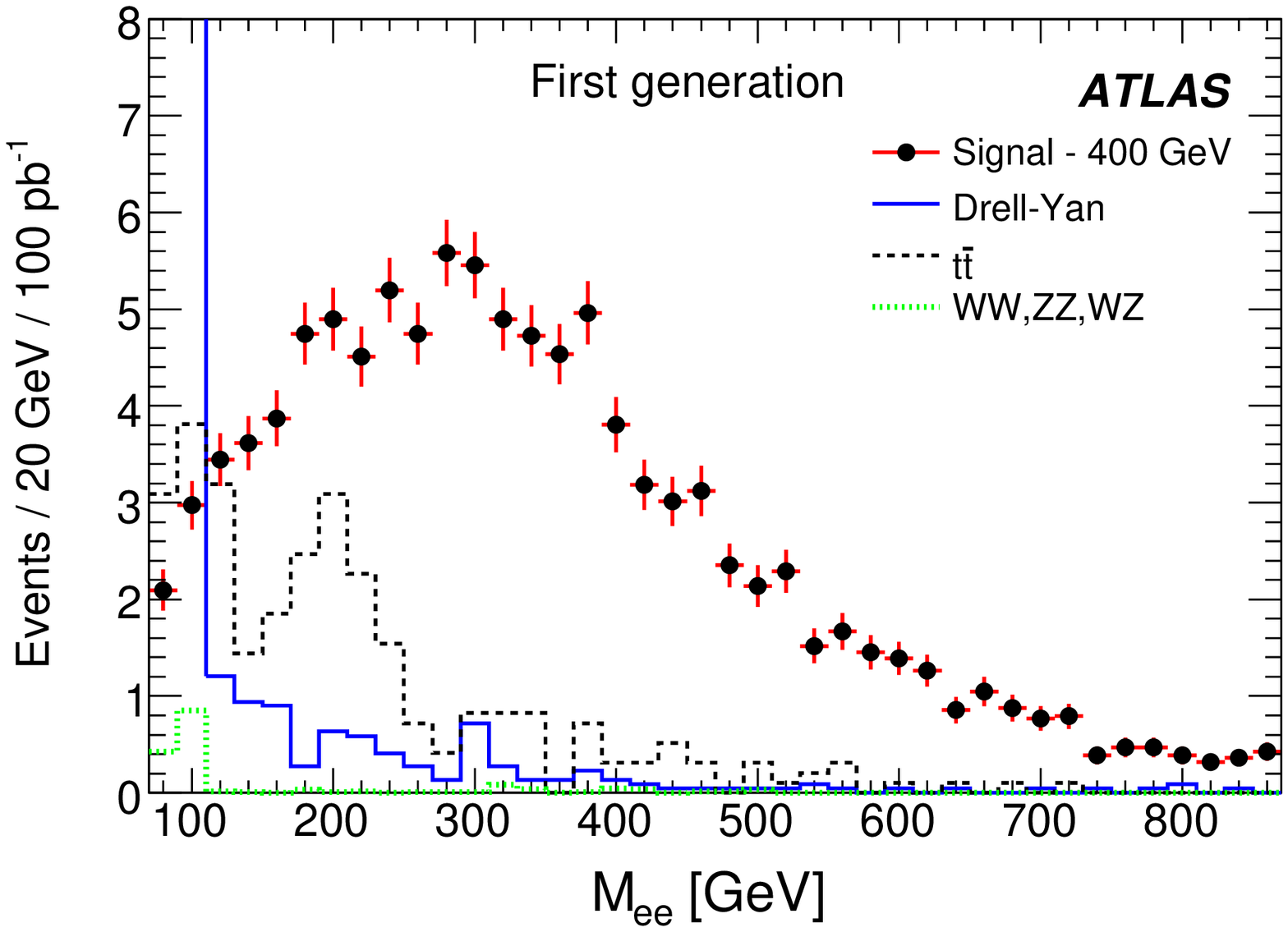}}
}
\caption{\small {
$S_{T}$ (left) and $M_{ee}$ of the selected electron pair after $S_{T}$ selection (right) in 1st gen.~leptoquark MC events (m$_{LQ} = $ 400 GeV). 
Both distributions are normalized to 100~$\pbarn$ of integrated $pp$ luminosity. 
}}
\label{Lq_ee_ST_Mee_cuts}
\end{figure}

The backgrounds are suppressed and the signal significance is enhanced by taking advantage of the fact that the final states in signal-like events have relatively large $\pt$. A scalar sum of transverse momenta of signal jets and lepton candidates, denoted by $\st$, helps in reducing the backgrounds while retaining most of the signal.  The other variable used to increase the signal significance is the invariant mass of the two leptons, $M_{ll}$. The distributions of these two variables are shown in Figure~\ref{Lq_ee_ST_Mee_cuts}.

After applying optimized selection on these two variables, $\st$ and $M_{ll}$, relative contribution of the background processes from $t\bar{t}$, $Z /DY$, diboson and multijet is 22\%, 7\%, 0.4\% and 18\%, respectively.
 
\begin{figure}[htbp]
\center{
{\includegraphics[width=2.4in]{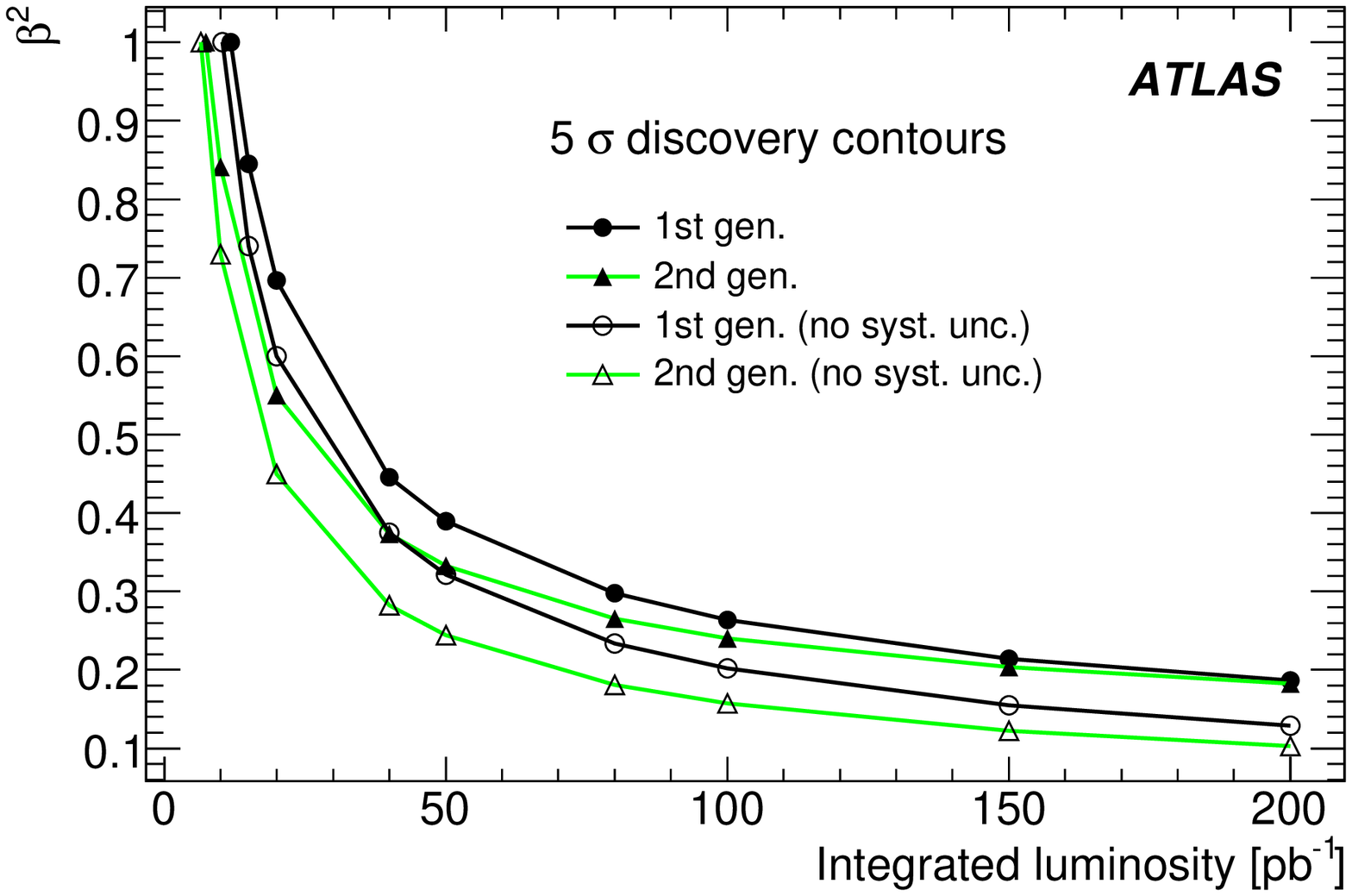}}
\hspace{0.5in}
{\includegraphics[width=2.4in]{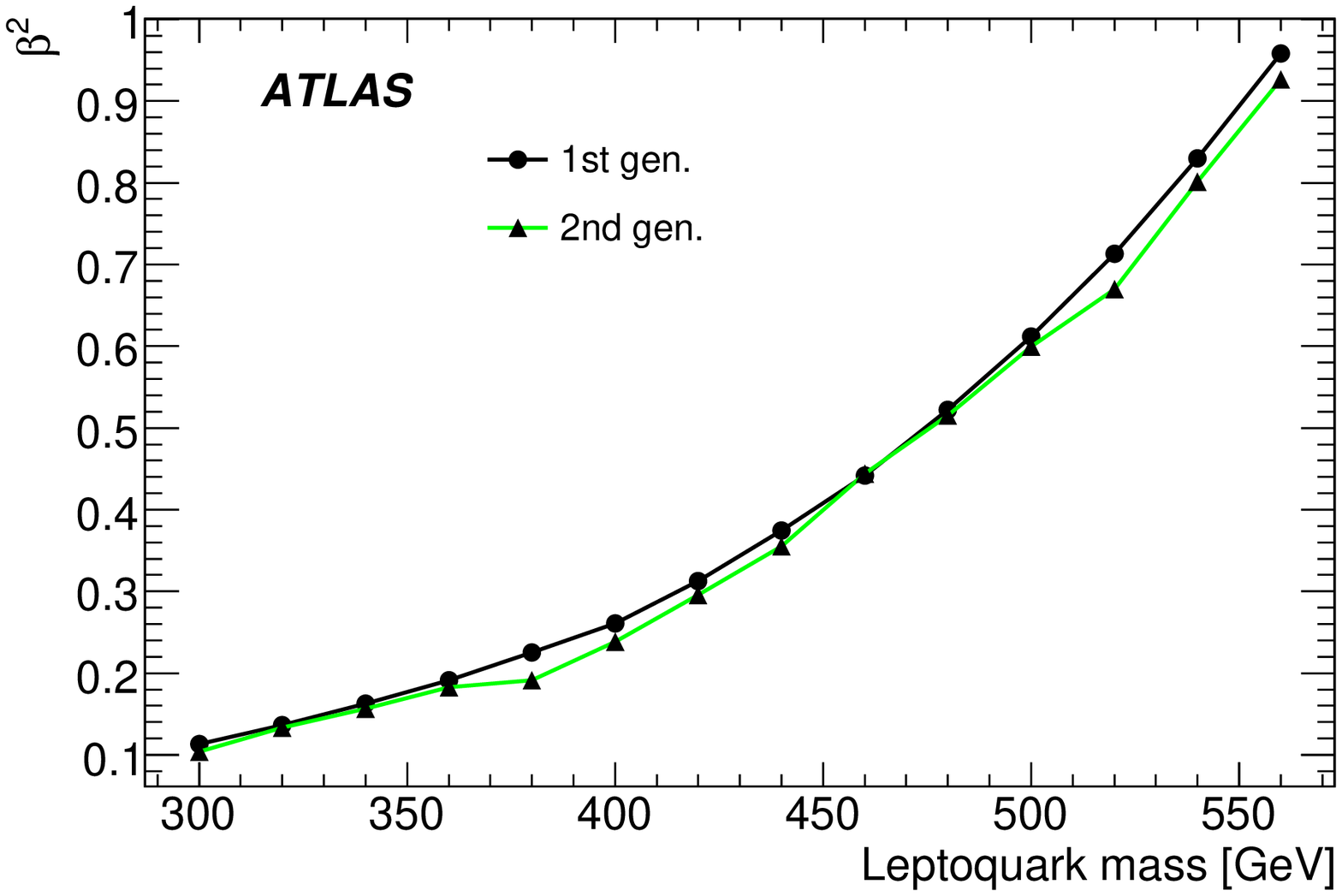}}
}
\caption{\small {
5$\sigma$ discovery potential for 1st and 2nd gen.~400~GeV  scalar leptoquark versus $\beta^2$, 
with and without background systematic uncertainty (left). Minimum $\beta^2$ of scalar leptoquark versus leptoquark mass at 100 $\pbarn$ of integrated $pp$ luminosity at 5$\sigma$ (background systematic uncertainty is included) (right). 
}}
\label{LQ_sensitivity}
\end{figure}

ATLAS's sensitivity to leptoquark signal with an integrated $pp$ luminosity of 100 $\pbarn$ is summarized in Figure~\ref{LQ_sensitivity}. The results include systematic uncertainty of 50\%. Trigger is based on a single high-$\pt$ lepton and is about 97\% efficient with respect to the offline reconstruction of the signal events. ATLAS is sensitive to leptoquark masses of about 565~GeV and 575~GeV for 1st and 2nd generations, respectively, at the given luminosity of 100 $\pbarn$ 

\section{EXCITED LEPTONS AT ATLAS}

\begin{figure}[htbp]
\center{
{\includegraphics[width=2.3in]{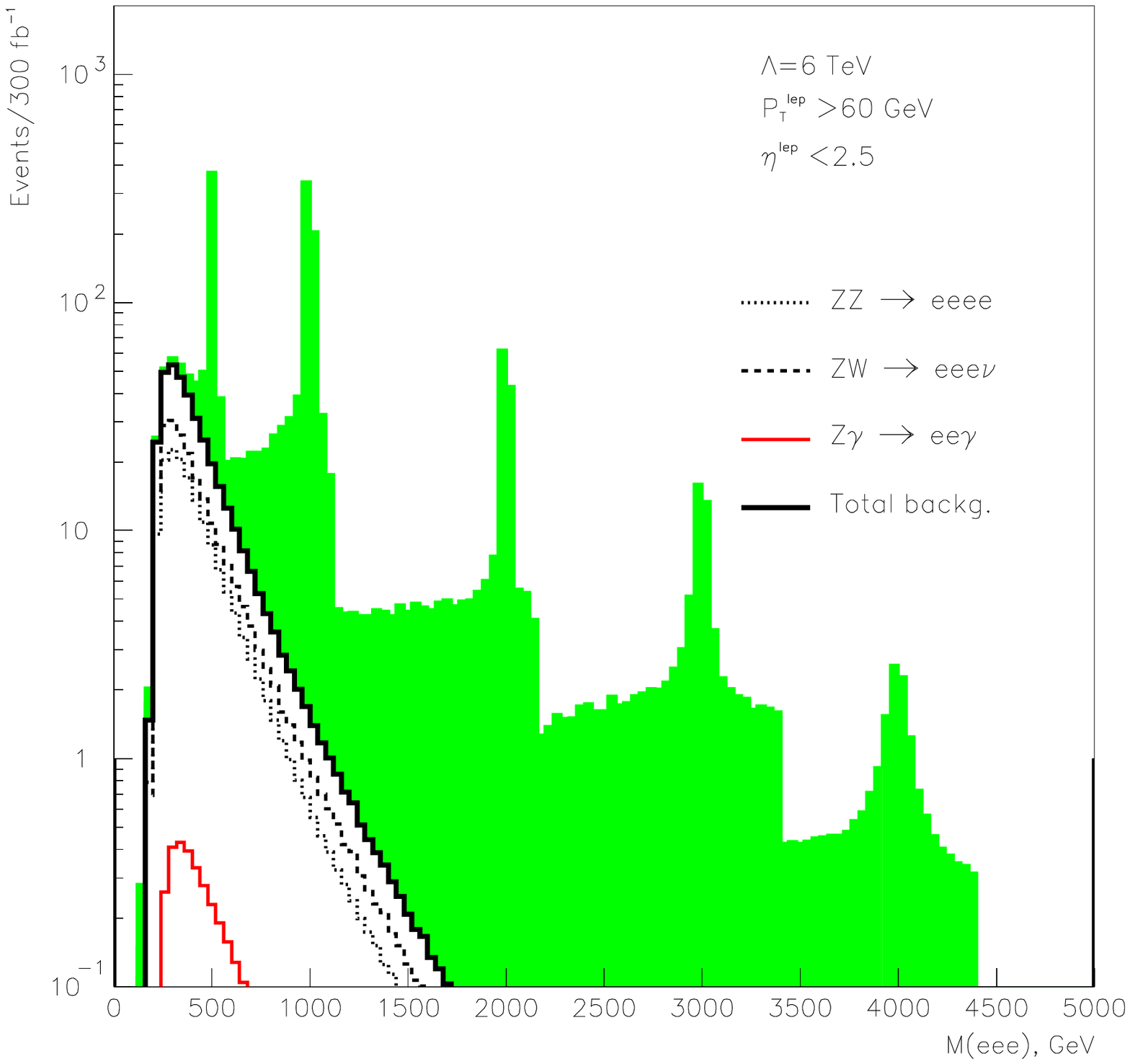}}
\hspace{0.5in}
{\includegraphics[width=2.3in]{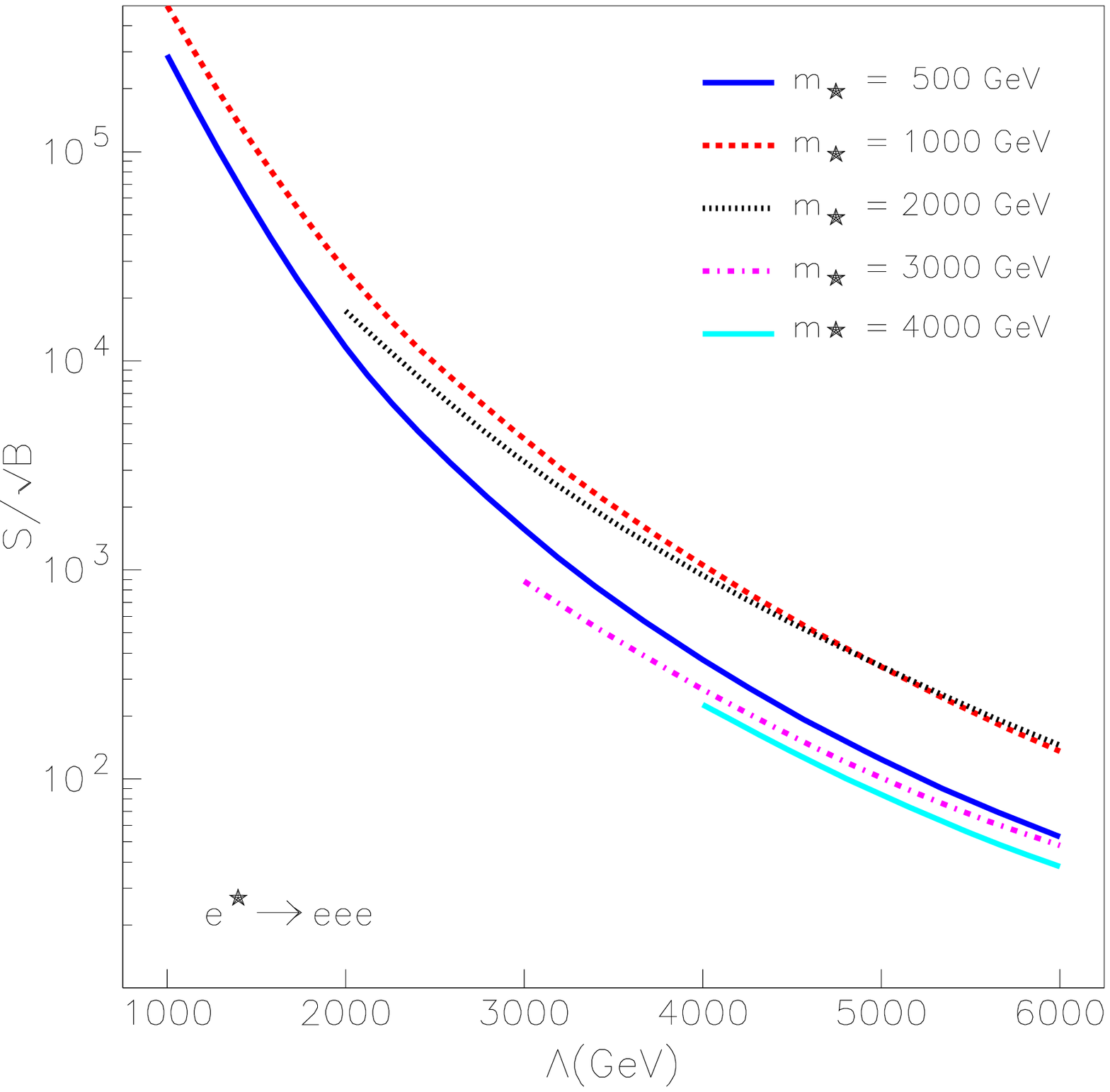}}
}
\caption{\small {
Invariant mass of excited electron candidate from three electrons (left). Excited electron signal significance shown for different mass hypothesis as a function of compositeness scale $\Lambda$ and (right). Both plots (ATLAS study) are for an integrated $pp$ luminosity of 300 $\fbarn$
}}
\label{excited_electron}
\end{figure}

Composite models\cite{Eichten:1983hw} predict the existence of quarks and leptons as bound states of three fermions or a fermion and a boson. The single production of excited electrons through contact interaction at the LHC has been studied in ATLAS \cite{Cakir:2002ATL, Cakir:2003} using MC samples simulated with {\sc Pythia} and fast simulation of the ATLAS detector \cite{Ritcher:1998}. Excited electrons can decay either via contact interaction or via an intermediate gauge boson. The final states studied include ee$\gamma$, eee, and ejj. Figure~\ref{excited_electron} shows the results for eee channel.
This study predicts that singly-produced excited electrons could be detected at the LHC with masses up to 3-4 TeV for $\Lambda=6$~TeV, where $\Lambda$ is the compositeness scale.

\section{TECHNICOLOR AT CMS}

Technicolor is an alternate mechanism of EWSB. It is based on strong symmetry breaking that introduces Goldstone bosons in the form of technipions that eventually provide masses to W$^{\pm}$ and Z$^0$ bosons. To be consistent with the observations explained by the SM, this hypothesis requires technicolor gauge couplings to evolve very slowly and predicts lightest technicolor resonances to have masses below a TeV. One of such light resonances studied at CMS is technirho($\rho_{TC}^{}$), predicted by Technicolor Straw Man Model (TCSM).

\begin{figure}[htbp]
\center{
{\includegraphics[width=2.5in]{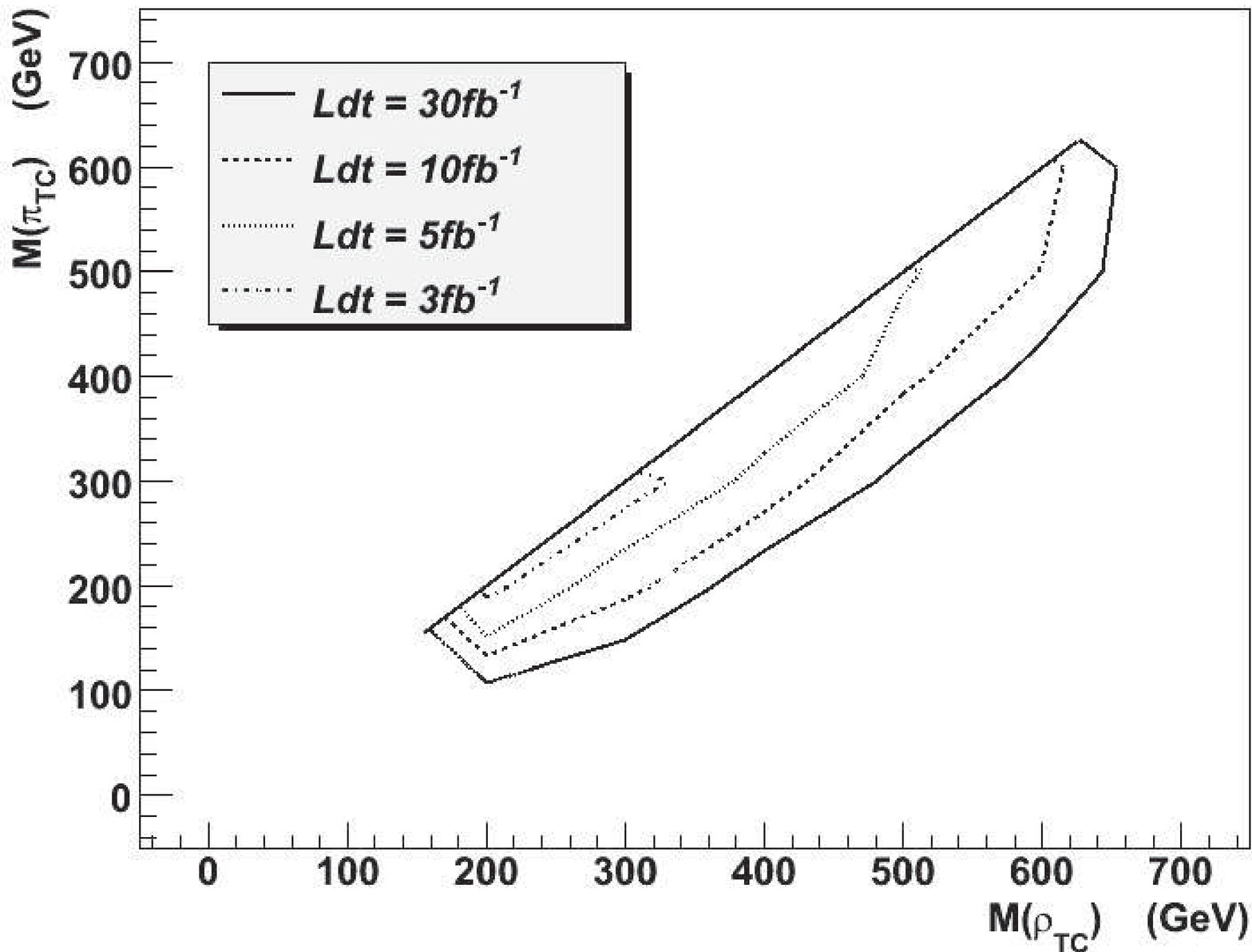}}
\hspace{0.5in}
{\includegraphics[width=2.5in]{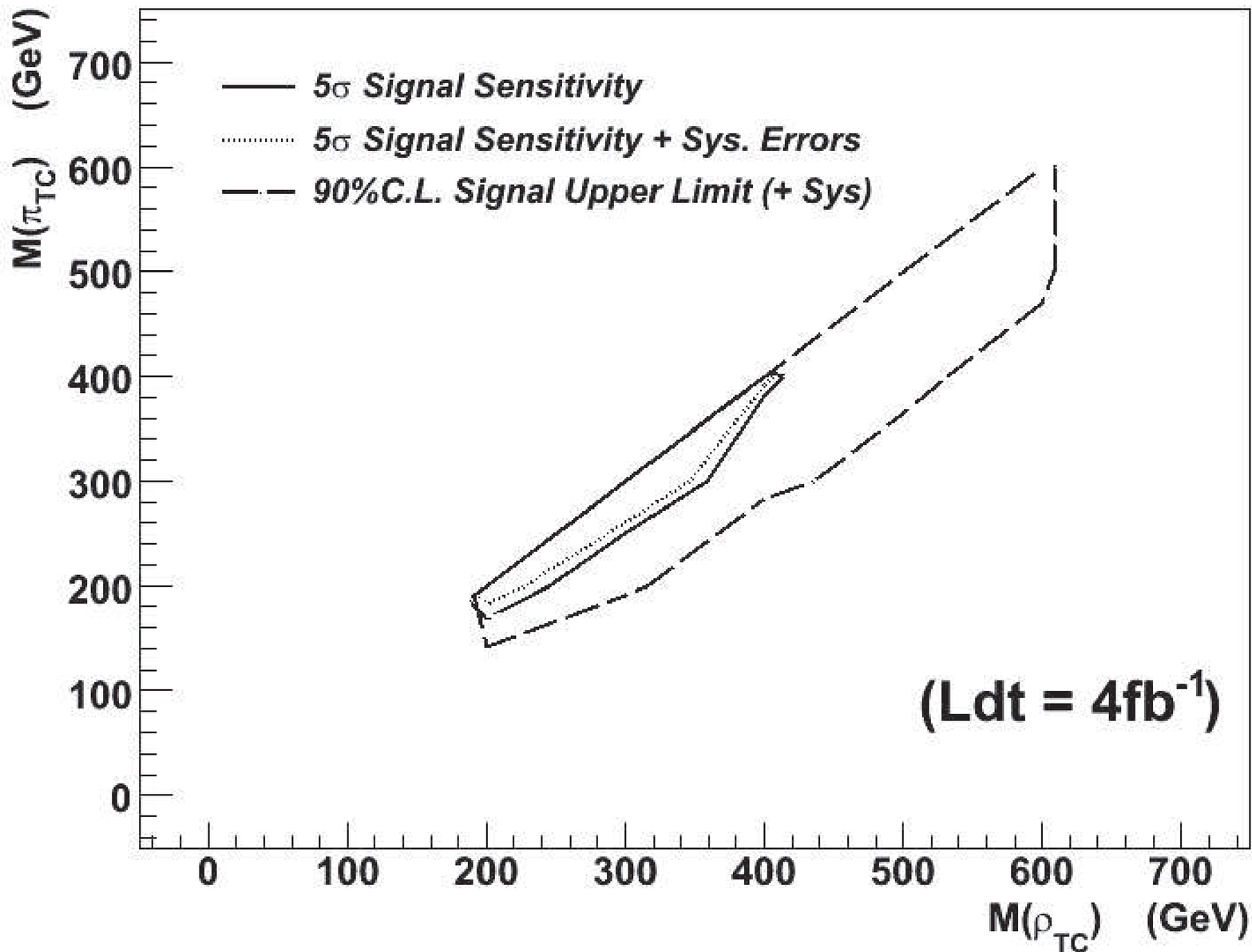}}
}
\caption{\small {
5$\sigma$ sensitivity curves (CMS study) for various integrated luminosities (left); sensitivity curve for 4 $\fbarn$ of integrated luminosity: the dotted (dashed) curve shows the sensitivity (90\% C.L. signal upper limit), systematic uncertainties included. 
}}
\label{TC_sensitivity}
\end{figure}

The $\rho_{TC}^{}$ analysis \cite{Kreuzer:2006} is performed in $3l+\met$~final state arising from decay channel $\rho_{TC}^{} \rightarrow W + Z$. The signal is simulated with {\sc Pythia} and fast simulation of the CMS detector \cite{CMSTDR1:2006}. Figure~\ref{TC_sensitivity} shows CMS sensitivity to Technicolor in technipion and technirho mass phase space for 4 $\fbarn$ of integrated $pp$ luminosity.

\end{document}